\input{aipcheck}

\documentclass[
    ,final            % use final for the camera ready runs
  ]
  {aipproc}

\layoutstyle{6x9}

\usepackage{graphicx}% eps fig, package for more complicated commands
\usepackage{amssymb} % various useful mathematical symbols
\usepackage{amsmath} % extended theorem environments
\usepackage{tabularx}
\usepackage{comment}
\usepackage{multirow}
\begin{document}

\title{Spin and diffractive physics with A Fixed-Target ExpeRiment at the LHC (AFTER@LHC)}


\author{C.~Lorc\'e}{
  address={IPNO, Universit\'e Paris-Sud, CNRS/IN2P3, F-91406, Orsay, France}
}

\author{M.~Anselmino}{
  address={INFN Sez. Torino, Via P. Giuria 1, I-10125, Torino, Italy}
}

\author{R.~Arnaldi}{
  address={INFN Sez. Torino, Via P. Giuria 1, I-10125, Torino, Italy}
}

\author{S.J.~Brodsky}{
  address={SLAC National Accelerator Laboratory, Stanford U., Stanford, CA 94309, USA}
}

\author{V.~Chambert}{
  address={IPNO, Universit\'e Paris-Sud, CNRS/IN2P3, F-91406, Orsay, France}
}

\author{J.P.~Didelez}{
  address={IPNO, Universit\'e Paris-Sud, CNRS/IN2P3, F-91406, Orsay, France}
}

\author{B.~Genolini}{
  address={IPNO, Universit\'e Paris-Sud, CNRS/IN2P3, F-91406, Orsay, France}
}

\author{E.G.~Ferreiro}{
  address={Departamento de F{\'\i}sica de Part{\'\i}culas, Univ. de Santiago de C., 15782 Santiago de C., Spain}
}

\author{F.~Fleuret}{
  address={Laboratoire Leprince Ringuet, \'Ecole Polytechnique, CNRS/IN2P3,  91128 Palaiseau, France}
}

\author{C.~Hadjidakis}{
  address={IPNO, Universit\'e Paris-Sud, CNRS/IN2P3, F-91406, Orsay, France}
}

\author{J.P.~Lansberg}{
  address={IPNO, Universit\'e Paris-Sud, CNRS/IN2P3, F-91406, Orsay, France}
}

\author{A.~Rakotozafindrabe}{
  address={IRFU/SPhN, CEA Saclay, 91191 Gif-sur-Yvette Cedex, France}
}

\author{P.~Rosier}{
  address={IPNO, Universit\'e Paris-Sud, CNRS/IN2P3, F-91406, Orsay, France}
}

\author{I.~Schienbein}{
  address={LPSC, Universit\'e Joseph Fourier, CNRS/IN2P3/INPG, F-38026 Grenoble, France}
}

\author{E.~Scomparin}{
  address={INFN Sez. Torino, Via P. Giuria 1, I-10125, Torino, Italy}
}

\author{U.I.~Uggerh\o j}{
  address={Department of Physics and Astronomy, University of Aarhus, Denmark}
}

\begin{abstract}
We report on the spin and diffractive physics at a future multi-purpose fixed-target experiment with proton and lead LHC beams extracted by a bent crystal. The LHC multi-TeV beams allow for the most energetic fixed-target experiments ever performed, opening new domains of particle and nuclear physics and complementing that of collider physics, in particular that of RHIC and the EIC projects. The luminosity achievable with AFTER using typical targets would surpass that of RHIC by more than 3 orders of magnitude. The fixed-target mode has the advantage to allow for measurements of single-spin asymmetries with polarized target as well as of single-diffractive processes in the target region.
\end{abstract}

\maketitle

\section{introduction}

Fixed-target experiment have been at the origin of numerous breakthrough in hadronic and nuclear physics. Thanks to the $7$ TeV proton and $2.76$ TeV lead LHC beams interacting on a (possibly polarized) fixed-target, many new opportunities open up and could be studied by a future multi-purpose experiment named AFTER, for ``A Fixed-Target ExpeRiment''. With the LHC proton beams, a center-of-mass energy close to $115$ GeV can be achieved for the first time in a fixed-target experiment. Moreover, these beams open a unique access to the (essentially unexplored) large negative-$x_F$ domain, allow for new investigations of the intrinsic heavy quark distributions, for a first precision measurement of the gluonic content of the neutron and, when the target is polarized, for a study of spin correlations including the Sivers effect.

The fixed-target mode has the advantage of reaching very high luminosities more easily than the collider mode, thanks to the density and length of the target, see Table \ref{tab:a}. Benefitting from nine months per year from the LHC proton runs, the production of heavy quarkonia, open heavy flavor hadrons and prompt photons in $pA$ collisions can be studied with unprecedented high statistics. Obviously, high-precision QCD measurements can also be carried out in $pp$ and $pd$ collisions with hydrogen and deuterium targets beside usual nuclear targets. More details and further physics opportunities can be found in Refs.~\cite{Brodsky:2012vg,Lansberg:2012kf}.

\begin{table}
\begin{tabular}{ccccccc}
\hline
\tablehead{1}{r}{b}{Beam}
  & \tablehead{1}{r}{b}{Target}
 & \tablehead{1}{r}{b}{Thickness (\rm{cm})}
  & \tablehead{1}{r}{b}{$\rho$ (\rm{g cm}$^{-3}$)}
  & \tablehead{1}{r}{b}{$A$}
  & \tablehead{1}{r}{b}{$\mathcal L$ ($\mu$\rm{b}$^{-1}$ \rm{s}$^{-1}$)} 
  & \tablehead{1}{r}{b}{$\int\mathcal L$ (\rm{pb}$^{-1}$ \rm{y}$^{-1}$)}  \\
\hline
$p$ & Solid H &10 & 0.088 & 1  & 260 & 2600\\
$p$ & Liquid H &100 & 0.068 & 1  & 2000 & 20000\\
$p$ & Liquid D & 100 & 0.16 & 2  & 2400 & 24000\\
$p$ & Pb & 1&11.35 & 207  & 16 & 160\\
\hline
Pb & Solid H &10& 0.088 & 1  & 0.11 & 0.11\\
Pb & Liquid H &100 & 0.068 & 1  & 0.8 & 0.8\\
Pb & Liquid D &100& 0.16 & 2  & 1 & 1\\
Pb & Pb & 1&11.35 & 207  & 0.007 & 0.007\\
\hline
\end{tabular}
\caption{Yearly ($10^7$s for $p$ and $10^6$s for Pb) integrated luminosities obtained with an extracted beam of $5\times 10^8$ $p^+$/s ($7$ TeV) and of $2\times 10^5$ Pb/s ($2.76$ TeV) for various targets.}
\label{tab:a}
\end{table}

\section{Studies for (semi-)exclusive physics at AFTER}

\begin{figure}[h]
  \begin{minipage}[b]{0.6\linewidth}
   \centering
   \includegraphics[height=3cm]{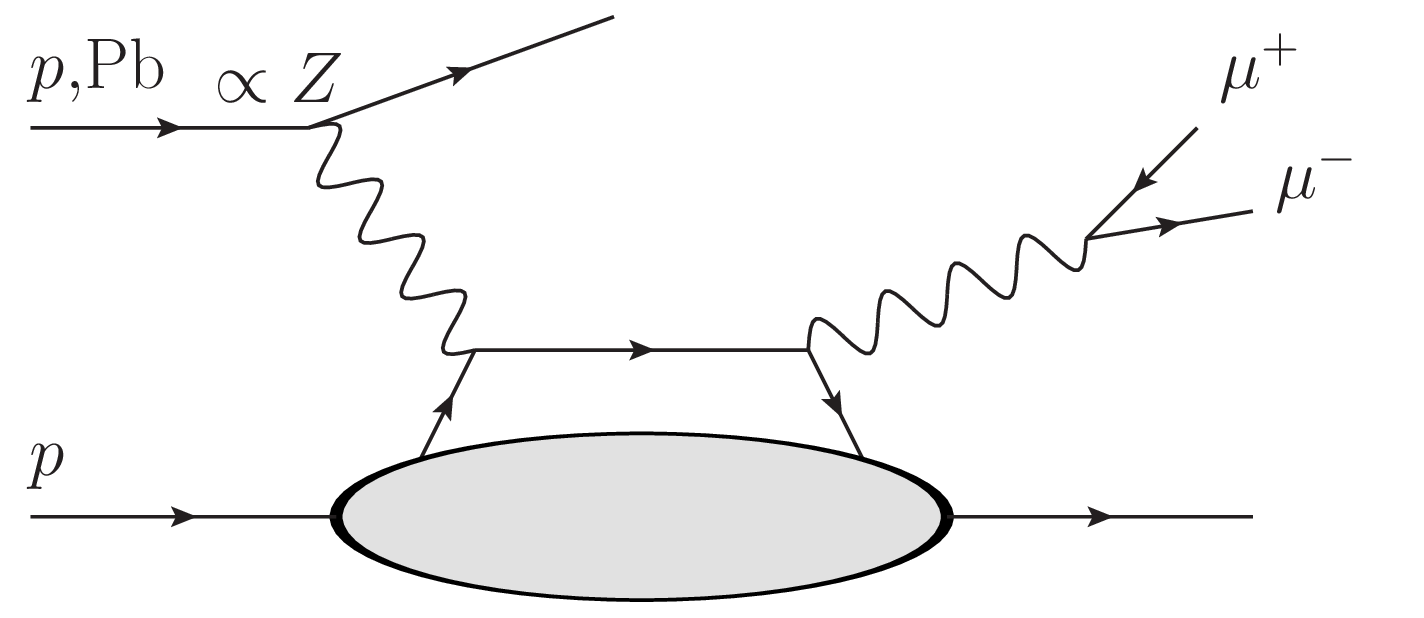}  
  \end{minipage}
  \begin{minipage}[b]{0.4\linewidth}
   \centering
   \includegraphics[height=3cm]{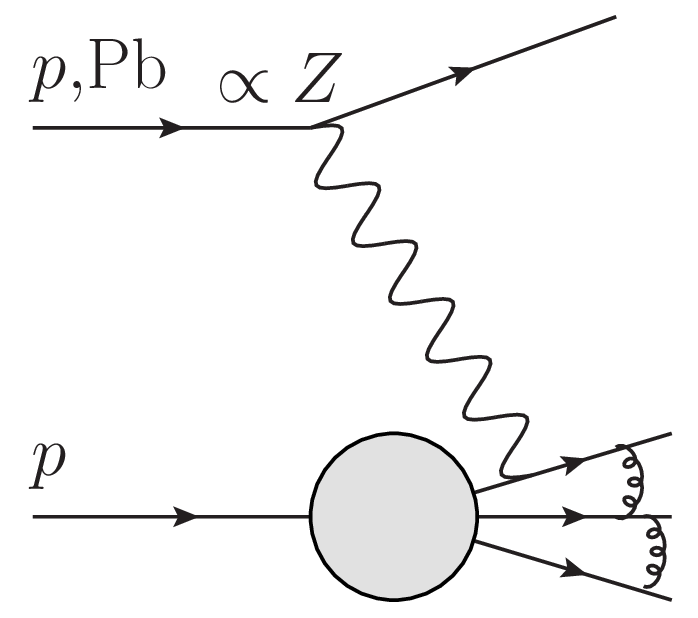}    
  \end{minipage}
 \caption{Left: Timelike DVCS gives access to the generalized parton distributions. Right: Diffractive dissociation of the proton into three jets gives access to the distribution amplitudes.}  \label{fig1}
\end{figure}

\noindent{\bf Ultra-peripheral collisions.} In Ultra-Peripheral Collisions (UPCs), two grazing nuclei (or even nucleons) interact electromagnetically and effectively turn heavy-ion colliders into photon colliders (for reviews see~\cite{Baur:2001jj,Bertulani:2005ru}). The highly boosted electromagnetic field of the nuclei allows for the production of hadronic systems \emph{via} the exchange of $2\gamma$, of $\gamma I\!\!P$, or of a single $\gamma$ accompanied by nucleon dissociation. The experimental proof of principle has been provided by RHIC~\cite{Adler:2002sc,Afanasiev:2009hy}. In the recent years, the study of the UPCs at the LHC has attracted a lot of interest \cite{Baltz:2007kq}. For instance, it was shown in~\cite{Strikman:2005yv} that the UPCs in $pA$ and $AA$ collisions would extend the coverage of HERA for the nuclear and gluon parton distribution functions (PDFs). Nevertheless, one of the main issue to face in $pp$ and $pA$ runs is the important pile-up, see~\emph{e.g.}~\cite{d'Enterria:2009er}. With the slow extraction of the LHC beams by means of a bent crystal, we have evaluated that the pile-up is absent for hydrogen targets. This offers many possibilities like the study of vector-meson elastic and inelastic photoproduction, \emph{e.g.} $pA \overset{\gamma}{\to} (X)~\psi(2S)+X~(A)$ which could not systematically be carried out by H1 and ZEUS, and timelike DVCS $pA \overset{\gamma}{\to} (p) ~\ell^+\ell^-~ (A)$ aiming at the extraction of generalized parton distributions (GPDs)~\cite{Pire:2008ea}, see figure \ref{fig1}. In particular, polarizing the target longitudinally would give acces to the polarized GPD $\tilde H$.
\newline

\noindent{\bf Heavy-hadron (diffractive) production at $x_F \to -1$.} Assuming that the constituents of a given intrinsic heavy-quark Fock state tend to have all the same rapidity, being that of the projectile or that of the target, an excess of quarkonia is expected in the forward ($x_F>0.1$) or backward ($x_F< -0.1$) region, respectively. This hypothesis can account for early $\Lambda_c^+$ data at $\sqrt{s}=62$ GeV~\cite{Basile:1981wh,Barger:1981sv}, and also possibly for the unexpected observations of leading $\Lambda_b$ at high $x_F$ at $\sqrt{s}=62$ GeV~\cite{Basile:1981nr}. Other hints for enhanced heavy-hadron production at large $x_F$ are the claimed production of double-charm $\Xi_{cc}^+$ baryons by SELEX~\cite{Mattson:2002vu,Ocherashvili:2004hi} (so far unseen in $e^+e^-$ reactions~\cite{Aubert:2006qw}) and the large-$x_F$ production of $J/\psi$ pairs at NA3~\cite{Badier:1982ae}, consistent with double-IC Fock states~\cite{Vogt:1995tf}. AFTER would be able to confirm or infirm such cross-section enhancement, for both charmed and beauty hadrons, and one could also look for the triply heavy baryons $\Omega^{++}_{ccc}$ and $\Omega^-_{bbb}$ undiscovered so far. Note also that, along the lines of Refs.~\cite{Chang:2011vx,Chang:2011du,An:2012kj}, it would also be particularly interesting to investigate the production of $\Lambda$ in the backward region as it may give some further evidences about the intrinsic light-quark sea.
\newline

\noindent{\bf Hard diffractive reactions.} Thanks to a full coverage of the target rapidity region, one could study the diffractive dissociation of the proton into three jets in PbH collisions (see figure~\ref{fig1}), thus measuring the three-quark valence light-front wavefunction of the projectile~\cite{Frankfurt:2002jq,Ivanov:2008ax}, in analogy to the E791 measurements of the diffractive dissociation of a pion jet~\cite{Aitala:2000hb,Ashery:2006zw}. In $pA$ collisions, the cross section should scale as $A^2 F^2_A(t)$, where $F_A(t)$ is the nuclear form factor~\cite{Frankfurt:2002jq}. Such nuclear dependence could be studied in PbH collisions by looking at mini-jets in the target region. Another opportunity is the possibility to test the pQCD color transparency~\cite{Brodsky:1988xz}: the prediction that there is no absorption of the initial state proton projectile in hard diffractive reactions. Such effect has been observed by E791~\cite{Aitala:2000hc} for $\pi$ projectiles but never for proton beams.
\newline

\noindent{\bf Very backward physics.} With deuterium target, it is possible to study hidden-color excitations of the deuteron~\cite{Matveev:1977xt}. There are five distinct color-singlet representations of the six-quark valence state of the deuteron. Only one of these, the $n-p$ state, is considered in the conventional nuclear physics (for a review see~\cite{Bergstrom:1979fp}), while all five Fock states mix by gluon exchange in pQCD. These novel ``hidden-color"  components~\cite{Brodsky:1983vf} can be studied at AFTER by probing parton distributions in inclusive reactions requiring high $x$ ($ \geq 1 $) and by studying the diffractive dissociation of the deuteron in its rapidity domain~\cite{Akhieser:1957zz} in Pb$d$ collisions.

\section{Summary}

A fixed-target facility (named AFTER) based on the multi-TeV proton or heavy ion beams at the LHC extracted by a bent crystal can provide a novel testing ground for QCD at unprecedented energies and momentum transfers. In this contribution, we briefly presented a short selection of opportunities for exclusive and semi-exclusive physics. Ultra-peripheral collisions allow for instance for studies of vector-meson elastic and inelastic production which could not be carried out by H1 and ZEUS, and of the generalized parton distributions \emph{via} timelike deeply virtual Compton scattering. Experiments at negative $x_F$ give the possibility to study intrinsic strangeness and charm in the proton, and give the possibility to look for the doubly and triply heavy baryons undiscovered so far. Hard diffractive reactions allows us to access directly the distribution amplitude of the proton in the diffractive dissociation of the proton into three jets, and to test the pQCD color transparency. Finally, in the very backward region, one can also look for hidden-color excitations of the deuteron, revealing the genuine six-quark structure beyond the conventional $p-n$ picture.

\end{document}